%% file: main.tex
\documentclass[twocolumn, english, prd,superscriptaddress,
aps, showpacs, showkeys, amsmath, amssymb, floatfix, nofootinbib]{revtex4-1}
\usepackage{multirow}
\usepackage[english]{babel}
\usepackage[utf8]{inputenc}
\usepackage[colorinlistoftodos, color=green!40, prependcaption]{todonotes}
\input{preamble}
\usepackage{bm}
\usepackage[flushleft]{threeparttable}
\usepackage{makecell,booktabs}
\usepackage{ulem}
\usepackage[pdftex, pdftitle={Article}, pdfauthor={Author}]{hyperref} 
\begin{document}
\title{Testing the consistency between cosmological data: the impact of spatial curvature and the dark energy EoS}

\author{Javier E. Gonzalez}
    \email{gonzalezsjavier@gmail.com}
  
\affiliation{Facultad de Ciencias e Ingeniería, Universidad de Manizales, 170002, Manizales, Colombia}  

\affiliation{Departamento de F\'{\i}sica, Universidade Federal do Rio Grande do Norte, 59078-970, Natal, RN, Brasil}

\author{Micol Benetti}
  \email{micol.benetti@unina.it}
\affiliation{Dipartimento di Fisica "E. Pancini", Universit\`{a} di Napoli  ``Federico II'', Complesso Universitario di Monte Sant'Angelo,
  Edificio G, Via Cinthia, I-80126, Napoli, Italy}
\affiliation{Istituto Nazionale di Fisica Nucleare (INFN) Sezione
  di Napoli, Complesso Universitario di Monte Sant'Angelo, Edificio G,
  Via Cinthia, I-80126, Napoli, Italy}
  \affiliation{Scuola Superiore Meridionale, Universit\`{a} di Napoli ``Federico II'', Largo San Marcellino 10, 80138 Napoli, Italy}

\author{Rodrigo von Marttens}
    \email{rodrigovonmarttens@gmail.com}
    \affiliation{Observat\'orio Nacional, 20921-400, Rio de Janeiro, RJ, Brasil}
    
\author{Jailson Alcaniz.}
    \email{alcaniz@on.br}
\affiliation{Observat\'orio Nacional, 20921-400, Rio de Janeiro, RJ, Brasil}

\date{\today} 

\begin{abstract}
The results of joint analyses of available cosmological data have motivated an important debate about a possible detection of a non-zero spatial curvature. If confirmed, such a result would imply a change in our present understanding of cosmic evolution with important theoretical and observational consequences.  In this paper we discuss the legitimacy of carrying out joint analyses with the currently available data sets and explore their implications for a non-flat universe and extensions of the standard cosmological model. 
We use a robust tension estimator to perform a quantitative analysis of the physical consistency between the latest data of Cosmic Microwave Background, type Ia supernovae, Baryonic Acoustic Oscillations and Cosmic Chronometers. We consider the flat and non-flat cases of the $\Lambda$CDM cosmology and of two dark energy models with a constant and varying  dark energy EoS parameter. The present study allows us to better understand if possible inconsistencies between these data sets are significant enough to make the results of their joint analyses misleading, as well as the actual dependence of such results with 
the spatial curvature and dark energy parameterizations.  

\end{abstract}


\maketitle

\section{Introduction}
Over the last two decades, the standard $\Lambda$ - Cold Dark Matter ($\Lambda$CDM) cosmology has been consolidated as the best description of cosmological observations \cite{Weinberg:2012es}. This model has shown great success in explaining the dynamics of the Universe up to first-order perturbations requiring only half a dozen parameters, namely: the physical baryon and CDM densities, the optical depth, the angular acoustic scale, the scalar spectral index, and the primordial curvature amplitude. 

In the concordance $\Lambda$CDM model, the  mechanism behind  the late-time cosmic acceleration is the cosmological constant  $\Lambda$,  characterized by a constant equation-of-state (EoS) parameter $w=-1$, and the Universe is assumed to be spatially 
flat. The latter assumption has been strongly supported by observational results obtained from joint analyses of many different cosmological observables  (see e.g. \cite{Aghanim:2018oex,Hinshaw:2012aka,Alam:2020sor,Scolnic:2017caz,Betoule:2014frx,Abbott:2018wog,Bonvin:2016crt,ACT,SPT}). From a statistical point of view, the addition of a non-zero curvature in such analyses does not seem to significantly improve the goodness-of-fit when compared with the flat scenario \cite{Efstathiou:2020wem} whereas from the theoretical point of view, the flatness of the Universe is well motivated by the simplest models of inflation \cite{PhysRevD.23.347,Linde:1982uu}, which suggests that the case for a zero spatial curvature is more than a mere simplification.

 An important aspect worth considering in this discussion concerns the actual consistency among the available cosmological data  and the need to introduce new degrees of freedom in the cosmological description to capture unknown features presented in the data. Recently, the latter possibility has been explored in several ways. In particular, given the latest results of the Planck Collaboration, a debate about a possible evidence for a positive curvature ($\Omega_k <0 $) in the power spectrum of the temperature fluctuations and polarization of the Cosmic Microwave Background (CMB) has emerged, which could imply a cosmic discordance between the current CMB data and other cosmological probes such as Baryon Acoustic Oscillations (BAO), Type Ia Supernovae (SNe Ia) and Cosmic Chronometers (CC) observations. 
 
 This apparent inconsistency has been pointed out by the Planck Collaboration \cite{Aghanim:2018oex} and interpreted by Di Valentino et al. \cite{DiValentino:2019qzk}, Handley \cite{Handley:2019tkm}, and Park \& Ratra \cite{Park:2017xbl} as a cosmic discordance in the framework of the $\Lambda$CDM cosmology. In Ref. \cite{DiValentino:2020hov} the authors explored a four-parameter extension of the standard model and presented cosmological constraints from CMB and SNe Ia data which exclude a flat universe at 99\% (CL).  A different interpretation of the results obtained from the currently available sets of observations was given by the Planck Collaboration \cite{Aghanim:2018oex} and more recently by Efstathiou \& Gratton \cite{Efstathiou:2020wem}. In the latter work, the authors performed a joint analysis of the main cosmological data to break the geometrical degeneracy of the parameters, showing that a flat universe is favoured. 

The tension level among the current data sets or, more specifically, if it allows performing a legitimate joint analysis is an important question not completely explored in these studies. Currently, the main inconsistency seem to arise between BAO and Planck data when the Plik likelihood is used, while the use of CamSpec likelihood provides a better consistency.  Furthermore, the geometrical degeneracy is partially broken when the CMB temperature and polarization data are combined with the CMB lensing information, which alleviates the CMB/BAO tension. It is also worth mentioning that the Planck evidence for a non-zero spatial curvature is not present in the the ACT Collaboration CMB data, which shows a good agreement with a flat universe \cite{ACT}. Another relevant question is whether these inconsistencies  also appear when one considers more general models e.g. a $\Lambda$CDM scenario with non-zero spatial curvature or even a flat universe dominated by a dark energy component with a more general EoS parameter ($w \neq -1$).

Currently, there is not a commonly accepted tension measure and a large number of estimators have been proposed in the literature e.g. the Bayes Evidence ratio ${\mathcal{T}}$ \cite{Verde:2013wza}, the Bayes Evidence ratio $R$ \cite{Marshall:2004zd}, the Suspiciousness \cite{Handley:2019wlz}, the parameter difference \cite{Battye:2014qga}, the Surprise \cite{Seehars:2014ora}, the Index of Inconsistency ($IOI$) \cite{Ishak2017a,Ishak2017b}, among others. 
The Bayes Evidence ratio is a commonly used tension estimator, however, like any quantity based on the evidence, it depends on a prior, which may affect the tension estimate. Since
our purpose is to use very large uninformative priors to allow data to provide the posterior distribution without biases from previous information, the Evidence Ratio is not a good choice. 

In our analysis we use the moment-based tension estimator $IOI$ to quantify the (in)consistency among the main cosmological data sets. {It is worth mentioning that the $IOI$, as well as other discordance estimators proposed in the literature, quantifies a mix of discrepancies caused by incomplete or incorrect models, systematic errors and data scatter. Therefore, in order to explore the consistency of different theoretical models from the compatibility of the main  cosmological data sets (rather than the data scatter), we adopt the novel approach presented in Ref \cite{Ishak2020}, the so-called level of physical inconsistency $\beta$ that quantifies only discrepancies of physical origin, i.e., due to wrong models or systematic errors.} In our analysis, we use the latest data of CMB, SNe Ia, BAO and CC,  and consider the flat and non-flat cases of the $\Lambda$CDM cosmology and of two dark energy models with a constant and varying  dark energy EoS parameter.  

This paper is organized as follows: in Sec. \ref{1} we present the primary data sets used in our analysis and the formalism adopted to estimate tensions among them. In Sec. \ref{2}, we show the confidence levels of the statistical analyses with the aim to expose graphically the data inconsistencies and, then, analyse the quantitative estimates of the tensions in Secs. IV and V. Finally, in Sec. VI, we present the main conclusions of this work.

\section{Observations and data tensions}
\label{1}

In this section, we shall present the observational data used in our analysis and the method  adopted to compute the consistency between data sets.

\subsection{Data}
\label{ssec:data}

We use the most commonly used data to perform parameter selection in the literature. They are the following:

\begin{itemize}
\item \textbf{CMB(Plik/CamSpec):} Cosmic Microwave Background measurements, through the Planck (2018) data~\cite{Aghanim:2019ame}, using ``TT,TE,EE+lowE" data by combination of temperature TT, polarization EE and their cross-correlation TE power spectra  over the range $\ell \in [30, 2508]$, the
low-$\ell$ temperature Commander likelihood, and the low-$\ell$ SimAll EE likelihood. Regarding the high-$\ell$, we analyze both the likelihood ``Plik" and ``CamSpec". The former is considered the more robust high-$\ell$ data of the Planck Collaboration, while the latter uses the same data of Plik with a variation on some key data and model choices (e.g., polarization mask and polarization efficiency). In polarization, the main differences are the use of a single mask to reduce the amount of computation required to calculate covariance; Galactic dust subtraction in polarization; effective calibration handling for TE and EE; the coaddition process; and the absence of polarized dust nuisance parameters. 

\item \textbf{Lens:} The CMB lensing reconstruction power spectrum composed by 9 correlated data points from the latest Planck satellite data release (2018)~\cite{Aghanim:2019ame,Aghanim:2018oex};

\item \textbf{Baryon Acoustic Oscillation (BAO)}: Following the analysis of the latest Planck release~\cite{Aghanim:2019ame}, we use the BAO measurements from 6dFGS~\cite{Beutler:2011hx}, SDSS-MGS~\cite{Ross:2014qpa}, and BOSS DR12~\cite{Alam:2016hwk} surveys;

\item \textbf{Type Ia SNe (Pantheon):} The Pantheon compilation~\cite{Scolnic:2017caz} contains 1048 SNe Ia in the redshift range $0.01< z <2.3$, which provides accurate measurements of the peak magnitudes in the rest frame of the B band $m_{B}$. The SNe Ia absolute magnitude $M$ is considered as a nuisance parameter.

\end{itemize}

\subsection{Discordance and physical inconsistency estimators}

As mentioned earlier, recent analyses combining the above data sets have led to different results on the constraints of the main cosmological parameters. 
In order to explore the consistency among these sets of observations, we adopt the Index of Inconsistency ($IOI$) to quantify possible tensions among them, as proposed in Refs.~\cite{Ishak2017a,Ishak2017b}. 

Given the parameter joint distributions of a model constrained by two different data sets, the $IOI$ is defined as
\begin{equation}
IOI\equiv \frac{1}{2}(\bm{\mu_1-\mu_2})(\bm{C_1+C_2})^{-1}(\bm{\mu_1-\mu_2})^{T},
\label{IOI}
\end{equation}
where $\bm{\mu_i}$ and $\bm{C_i}$ correspond to the mean parameter vector and its covariance matrix, respectively, constrained by the $i$-th data set. This tension estimator considers the difference between the parameter mean vectors of the posteriors and also the effective size of the parameter space encoded in the covariance matrix. 

It is important to highlight that the tension measure does not depend only on the data sets, but also on the model adopted in the analysis.  Other crucial point is that the analysis of tensions must not be carried out parameter by parameter (marginalizing over the other ones), but a joint calculation that takes into account the complete set of parameters and their correlations, i.e., the  joint posterior distribution of the model parameters. The main reason for the joint approach is that the tension analysis considering marginalized distributions of separated parameters can hide disagreements due to the parameter correlations, as shown in Ref.~\cite{Ishak2017a,Lemos:2020jry}.

\begin{table*}[]
\begin{tabular}{cccc} \hline\hline
\multicolumn{4}{c}{\textbf{}} \\
\multicolumn{4}{c}{\textbf{No evidence of physical inconsistency exceeding the threshold}}     \\
\multicolumn{4}{c}{$P(\beta>1|\sqrt{2IOI})<(1-\alpha)$ (Ranking number: 1)}                     \\ \hline \hline
\multicolumn{4}{c}{\textbf{}}                                                                   \\
\multicolumn{4}{c}{\textbf{Evidence of physical inconsistency exceeding the threshold}}  
                \\
\multicolumn{4}{c}{$P(\beta>1|\sqrt{2IOI})>(1-\alpha)$}                                                       \\ \hline
\multicolumn{1}{c|}{\textbf{Single $\beta$ value}}  & \multicolumn{1}{c|}{\textbf{Level inconsistency}} & \multicolumn{1}{c|}{\textbf{\begin{tabular}[c]{@{}c@{}}$\beta$ range\\ $[\beta_{lower},\beta_{upper})$\end{tabular}}}                                                                     & \textbf{\begin{tabular}[c]{@{}c@{}}Ranking inconsistency\\ (Ranking number)\end{tabular}} \\ \hline
\multicolumn{1}{c|}{\multirow{2}{*}{$\beta$<3.5}}   & \multicolumn{1}{c|}{\multirow{2}{*}{Substantial}}    & \multicolumn{1}{c|}{\begin{tabular}[c]{@{}c@{}}Inside the \\ substantial level\end{tabular}}                    & \begin{tabular}[c]{@{}c@{}}Substantial\\ (2)\end{tabular}                                    \\ \cline{3-4} 
\multicolumn{1}{c|}{}                               & \multicolumn{1}{c|}{}                             & \multicolumn{1}{c|}{\begin{tabular}[c]{@{}c@{}}Spans substantial \\ and strong level\end{tabular}}              & \begin{tabular}[c]{@{}c@{}}Substantial-to-strong\\ (3)\end{tabular}                          \\ \hline
\multicolumn{1}{c|}{\multirow{2}{*}{3.5<$\beta$<5}} & \multicolumn{1}{c|}{\multirow{2}{*}{Strong}}      & \multicolumn{1}{c|}{\begin{tabular}[c]{@{}c@{}}Spans substantial, strong \\ and very strong level\end{tabular}} & \begin{tabular}[c]{@{}c@{}}Strong\\ (4)\end{tabular}                                      \\ \cline{3-4} 
\multicolumn{1}{c|}{}                               & \multicolumn{1}{c|}{}                             & \multicolumn{1}{c|}{\begin{tabular}[c]{@{}c@{}}Spans strong and \\ very strong level\end{tabular}}           & \begin{tabular}[c]{@{}c@{}}Strong-to-very\\ strong (5)\end{tabular}                       \\ \hline
\multicolumn{1}{c|}{5>$\beta$}                      & \multicolumn{1}{c|}{Very strong}                  & \multicolumn{1}{c|}{Inside the strong level}                                                                 & \begin{tabular}[c]{@{}c@{}}Very strong\\ (6)\end{tabular}                                 \\ \hline
\end{tabular}
\caption{Guiding interpretation of the level of physical inconsistency. The inconsistency levels are motivated by the interpretation of the 3$\sigma$, 4$\sigma$ and 5$\sigma$ tension in 1D distribution considering a specific value of $\beta$. The ranking of inconsistency is a more suitable scale that takes into account that $\beta$ is characterized by a range of values rather than a single number, i.e., the $\beta$ distribution.   The first rank corresponds to probability of $\beta>1$ lower than the threshold of significant tension, $P=1-\alpha=0.85$ \cite{Ishak2020}.}
\label{scale}
\end{table*}

As any other discordance estimator in literature, the $IOI$ measures a combination of physical inconsistency,  which may be caused by incomplete/incorrect model or systematic errors, and data scatter. In order to quantify uniquely the level of physical inconsistency, $\beta$,  we adopt the novel approach proposed in Ref \cite{Ishak2020}. The main idea of this novel approach is to calculate the  conditional probability of the physical inconsistency, given a discordance estimator value, and 
to infer from this distribution  the discordance estimators values that probabilistically imply levels of physical inconsistency that cannot be neglected. For this purpose it is used the Bayes' theorem,

\begin{equation}
P(\beta|\sqrt{2IOI})=\frac{P(\sqrt{2IOI}|\beta)P(\beta)}{P(\sqrt{2IOI})},
\end{equation}
with its usual interpretation. 

The first step is to estimate if the discordance measure  (in our case the $IOI$)  implies 
that the probability $P(\beta>1|\sqrt{2IOI}$) exceeds a specific level $P$, i.e.,
\begin{equation}
\int_{1}^\infty P(\beta|\sqrt{2IOI}) d\beta>P.
\label{threshold}
\end{equation}
The criterion adopted to define the probability threshold is the $\alpha$ significance level with $\alpha=0.15$ ($P=1-\alpha$). This choice is due to the fact that in an one-parameter distribution the $\alpha$ value corresponds to 2$\sigma$ confidence level (See Table 1 of Ref. \cite{Ishak2020}).  The second step is to obtain the most representative value of $\beta$ and its credible interval. For this purpose it is used the median statistics, with the median value of $\beta$ ($\beta_{med}$) defined as
\begin{equation}
\int_{0}^{\beta_{med}}P(\beta|\sqrt{2IOI}) d\beta=0.5\;,
\end{equation}
and its 68\%-percentile lower and upper limits 
\begin{equation}
 \int^{\beta_{lower}(\infty)}_{0(\beta_{upper)}}P(\beta|\sqrt{2IOI}) d\beta=0.16.
\end{equation}The next step is to choose an adequate scale to interpret the $\beta$ results. The scale proposed in Ref. \cite{Ishak2020} is motivated by the  classification performed in Ref. \cite{Verde:2019ivm} based on the $n$-$\sigma$ and it is shown in Table \ref{scale}\footnote{ An available code to quantify $\beta$ can be found in https://github.com/WeikangLin/IOI. }. See also Ref. \cite{Ishak2020} for a complete discussion about the level of physical inconsistency.\footnote{ As  noted in Ref. \cite{Ishak2017b}, the original term "moderate tension" should not be interpreted as an irrelevant inconsistency, mainly if the value is close to the upper limit. In order to avoid an underestimation of this tension level, we replace the original term "moderate" to "substantial".}

In addition to the $IOI$, it is also useful to define a quantity that allows identifying  outliers in a composition of multiple data sets. For this purpose, we use the Outlier Index (${\mathcal{O}}_j$)  that consists in the calculation of the $IOI$ between the $j$-th data set, the one considered as a possible outlier, and a joint multiple data sets, which must not include the $j$-th data set. Thus, the ${\mathcal{O}}_j$ value is obtained by considering in Eq. (\ref{IOI})  the mean and covariance matrix of the parameters constrained by the $j$-th data set and the ones constrained by the joint analysis of the other data sets (without considering the $j$-th data set) and subtracting the term  $(N_p-1)/2$, where $N_p$ is the number of common parameters of the mean vectors and matrices.
The interpretation of the outlier index is made in relative terms by comparing the ${\mathcal{O}}_j$ value of each data set. An outlier data is identified when its own ${\mathcal{O}}_j$ value  is significantly higher than the other ones \cite{Lin2019}. 

\begin{figure*}[t!]
\centering
\includegraphics[width=0.70\textwidth]{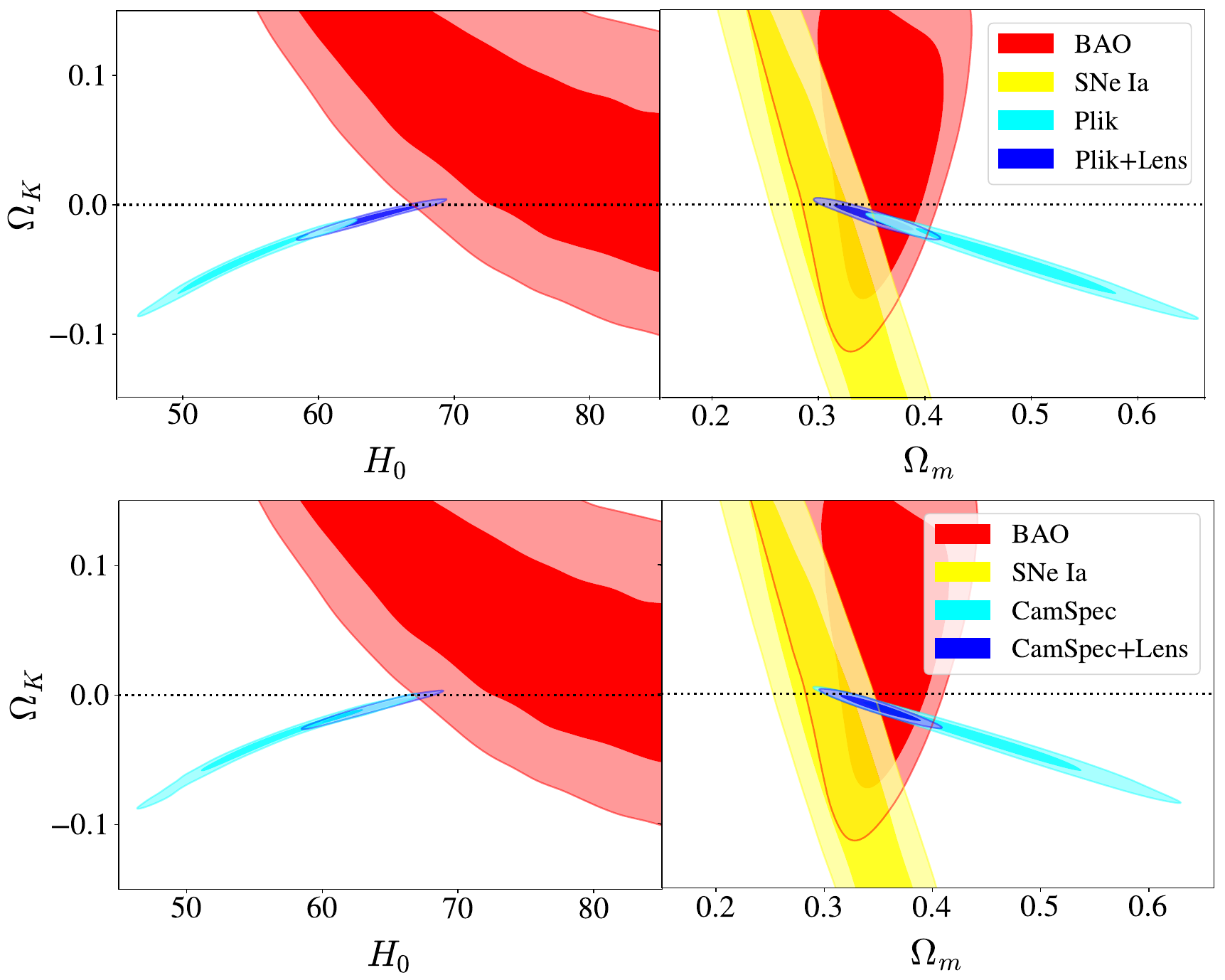}
\caption{Constraints on the $H_0-\Omega_k$ and $\Omega_m-\Omega_k$ planes from SNe, BAO and CMB (+Lens) data considering the $o\Lambda$CDM model. In the upper panels we present the confidence contours for CMB (+Lens) data using the Plik likelihood whereas in the lower panels we use the CampSpec CMB (+Lens) likelihood.}
\label{fig:LCDM} 
\end{figure*}

\section{Model Analysis }
\label{2}

We shall now perform the statistical analysis that will be used to compute the consistency between the data sets and to identify possible outliers. Here, in order to give a first qualitative idea of the problem, we show some contour planes to illustrate our main results discussed in Sec.~\ref{3}. However, we emphasize that a correct consistency test cannot be made regarding only 2D contours separately, but must take into account all the parameter space simultaneously.

The models we consider are the following:
\begin{itemize}
    \item \textbf{$\Lambda$CDM($+\Omega_{k}$):} The standard cosmological model 
    { with zero and non-zero spatial curvature as a free parameter, which we denote as ``$\Lambda$CDM'' and ``$o\Lambda$CDM'', respectively.} 
    
    \item\textbf{$w$CDM($+\Omega_{k}$):} A simple one-parameter extension of the previous model, where the DE component is no longer described by the cosmological constant, but by a dark component  with constant EoS $w$. As well as in the first case, we also consider the model with spatial curvature as a free parameter. From now on, the flat and non-flat model of this kind are denoted by ``$w$CDM'' and ``$ow$CDM'', respectively;
    
    \item\textbf{CPL($+\Omega_{k}$):} A two-parameter extension of the $o\Lambda$CDM, where the DE component is described by the CPL parameterization, i.e., 
    a time-dependent DE EoS given by $w\left(a\right)=w_{0}+w_{a}\left(1+a\right)$ \cite{Chevallier2000,Linder2002}. This parameterization is of particular interest because it is widely used for parameter selection~\cite{Aghanim:2019ame} and forecasts \cite{Resco:2019xve} (see also \cite{Barboza2008,Barboza2009,Lazkoz2010,Jassal2004,Adak2012} for other DE EoS parameterizations). From now on, the flat and non-flat model of this kind is denoted by ``CPL'' and ``$o$CPL'', respectively.
\end{itemize}

We divide our analysis in two parts. First, in order to compute the physical consistency among the individual probes, we perform a parameter estimation analysis with the individual probes presented in Sec.~\ref{ssec:data}. The second analysis performed is for the outlier diagnostic. In this case we combine the data sets in pairs to be compared with the remaining data set. Even though CMB and CMB+Lens are considered single data sets separately, they are not combined. Thus, the possible groups are: CMB+BAO, (CMB+Lens)+BAO, CMB+SNe Ia, (CMB+Lens)+SNe Ia, and BAO+SNe Ia. The last group, BAO+SNe Ia, is compared with CMB and with CMB+Lens. In all cases, CMB stands for both Plik and CamSpec likelihoods.

\begin{figure*}[t!]
\centering
\includegraphics[width=0.96\textwidth]{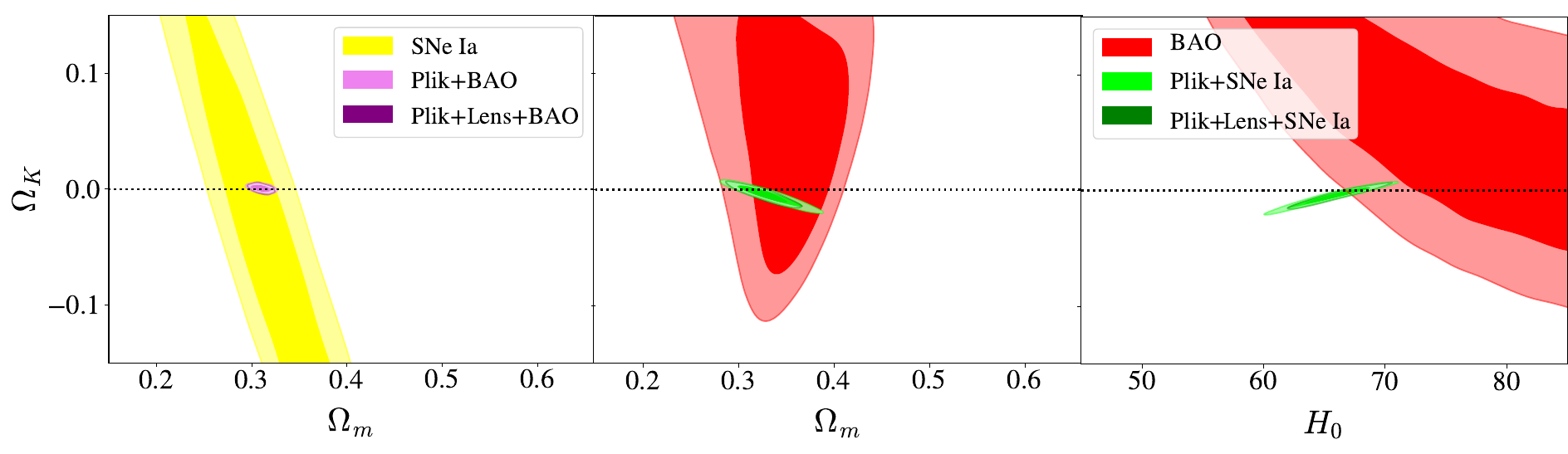}

\caption{The $1\sigma$ and $2\sigma$  constraints on the $H_0-\Omega_k$ and $\Omega_m-\Omega_k$ planes from single data sets. }
\label{fig:LCDMoutliers1} 
\end{figure*}

\begin{figure*}[t!]
\centering
\includegraphics[width=0.70\textwidth]{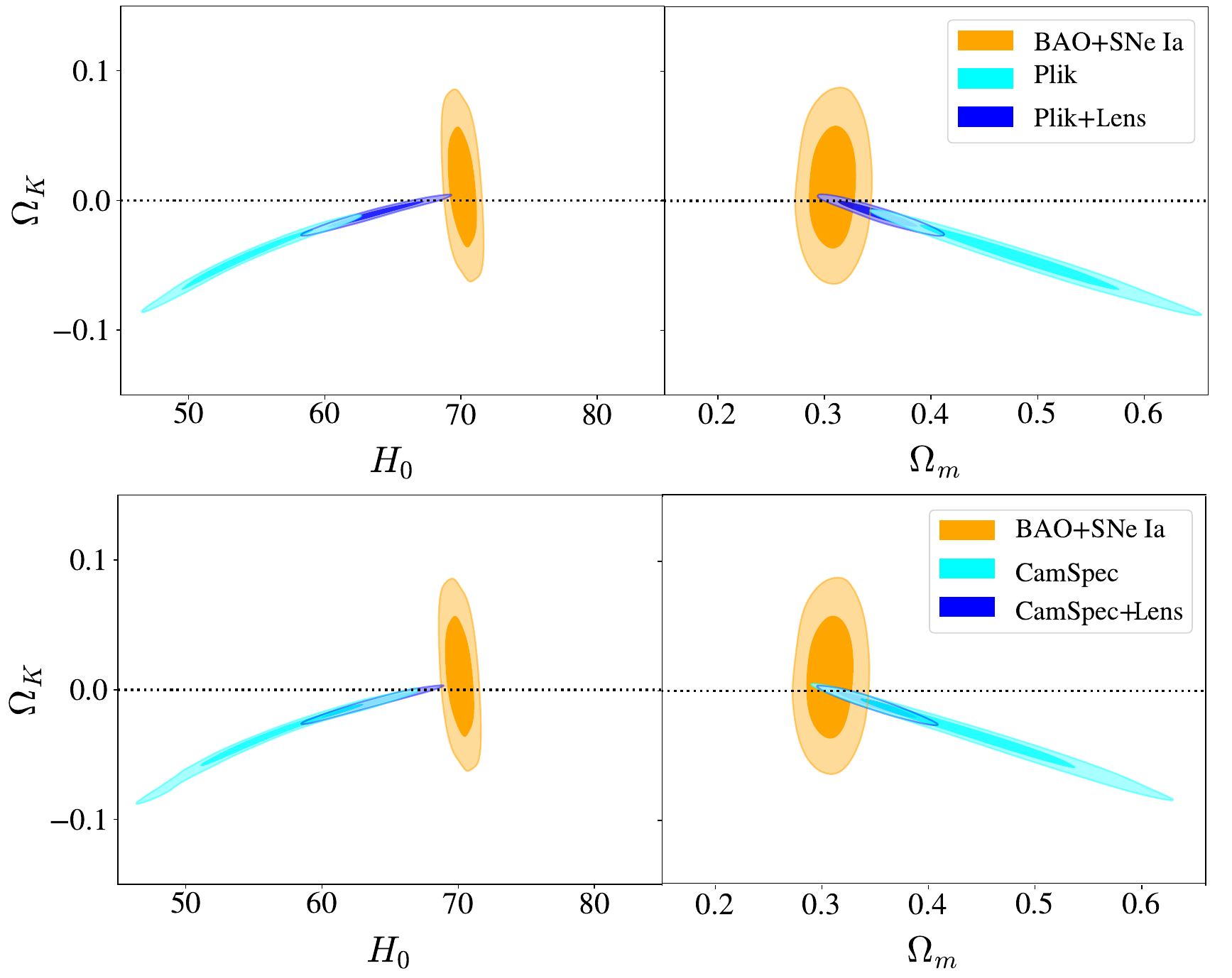}
\caption{Current $1\sigma$ and $2\sigma$  constraints on the $H_0-\Omega_k$ and $\Omega_m-\Omega_k$ planes from combination of data sets. The search for  outliers is performed for CMB data (Plik and CamSpec likelihoods).}
\label{fig:LCDMoutliers2} 
\end{figure*}

In Fig. \ref{fig:LCDM}, we show the results of the $H_0-\Omega_k$ and $\Omega_m-\Omega_k$ planes for the $o\Lambda$CDM model using the independent data sets separately. Since type Ia SNe cannot constraint $H_{0}$, all panels that contain this quantity do not show results from SNe Ia. 
Note that 
both BAO and SNe data seem to be in tension with CMB (without Lens) data, being more evident in the $H_0-\Omega_k$ plane for BAO data and in the $\Omega_m-\Omega_k$ for SNe Ia observations. In the latter case, the tension arises because of the difference in the $\Omega_m$ estimates, as SNe data do not provide any information about $H_0$ and do not constrain effectively the curvature. This disagreement is reduced if one uses the CamSpec likelihood instead of the Plik likelihood, the primary Planck Collaboration likelihood. Clearly, the inclusion of the CMB lensing breaks the geometrical degeneracy for both CMB likelihoods, as pointed out in \cite{Aghanim:2018oex},  and the principal source of tension between BAO and CMB+Lens data remains in the $H_0$ estimate.  As shown in the right panels of Fig. \ref{fig:LCDM}, the inclusion of CMB lensing data to  CMB data reduces considerably the SNe/CMB tension, mainly when it is used the Plik likelihood, which will be quantified in the next section. 
On the other hand, BAO and SNe Ia data show a good agreement. As it was discussed in previous works, the CamSpec allows better compatibility with a flat universe than the Plik likelihood, whilst the CMB+Lens, SNe, and BAO data sets are consistent with $\Omega_k = 0$ \cite{Efstathiou:2020wem,Handley:2019wlz}.

\begin{figure*}[t]
\centering
\includegraphics[width=1.0\textwidth]{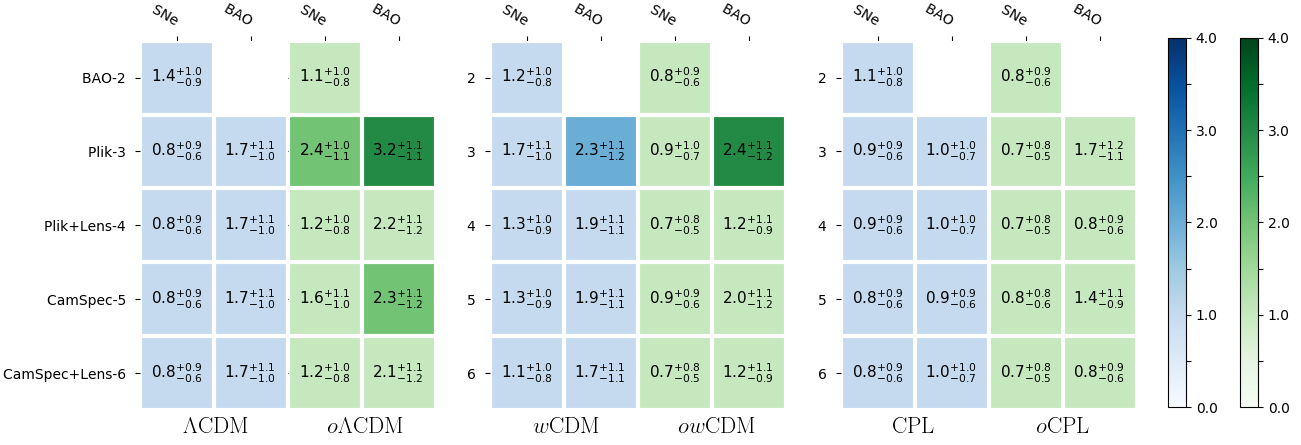}
\caption{Values of the level of physical tension $\beta$ between independent data sets assuming a flat universe (blue) and a universe with curvature (green). 
The \textit{left, middle} and \textit{right} panels correspond to the matrix tensions in the $\Lambda$CDM ($o\Lambda$CDM), $w$CDM ($ow$CDM) and CPL ($o$CPL) models, respectively. The color scale represents the ranking number introduced in Table \ref{scale}. Note that the highest inconsistency rank obtained is substantial-to-strong (3).}
\label{fig:tensions} 
\end{figure*}

In order to identify possible outlier data sets, we show in Fig. \ref{fig:LCDMoutliers1} and \ref{fig:LCDMoutliers2} the constraints from a single observable and from the joint analyses of the  remaining data sets, respectively. In Fig. \ref{fig:LCDMoutliers1}, we show the results for the outlier analysis considering the CMB(+Lens) combined with BAO and SNe Ia. We choose to show only the Plik likelihood case because the results using both CMB likelihoods are very similar. As one may see, the SNe constraints are in a very good agreement with the CMB(+Lens)+BAO constraints. In the second and third panels, we explore the possibility of the  BAO data being an outlier. In the $\Omega_m-\Omega_k$ plane we note a good compatibility between BAO and CMB(+Lens)+SNe whereas in the $H_0-\Omega_k$ plane a certain difference in the contours is evident. In these cases the addition of SNe Ia or BAO observations to the CMB(+Lens) data is enough to break the parameter degeneracy, making the CMB lensing not very relevant.

In Fig.~\ref{fig:LCDMoutliers2}, we present the SNe+BAO and CMB constraints. In both planes, $\Omega_m-\Omega_k$ and $H_0-\Omega_k$, it is possible to visualize  the strong inconsistency between  the late and early-time data when the Plik CMB likelihood is considered. However,  this inconsistency is considerably reduced if the CMB data is analysed with the CamSpec likelihood or if the CMB lensing is also included in the analysis (CMB+Lens).

We also explore the observational constraints on cosmological parameters in the framework of the $ow$CDM model. Contrary to the results for the $o\Lambda$CDM model,  the analysis for the $ow$CDM model shows that a  non-null curvature  is not confirmed  at 2$\sigma$  confidence level using only CMB data, and when lensing data are also considered, the spatial geometry is  compatible with a flat universe at 1$\sigma$ confidence level.  However, for Plik CMB likelihood the curvature parameter $\Omega_k=0$  needs high $H_0$ values and an EoS crossing the phantom line, while for the CamSpec likelihood we obtain more compatible results with the standard cosmology. Apparently, the $ow$CDM model can alleviate the inconsistencies between all data sets independently of the CMB likelihood. However, such result needs to be confirmed by a consistency estimator as the marginalized  distributions can hide possible tensions, which we perform in the next section. In the case of the $o$CPL model, the constraints are not very restrictive  when  a single observable is considered and its analysis of tensions from the marginalized confidence contours is not very useful.

\begin{figure*}[t]
\centering
\includegraphics[width=0.8\textwidth]{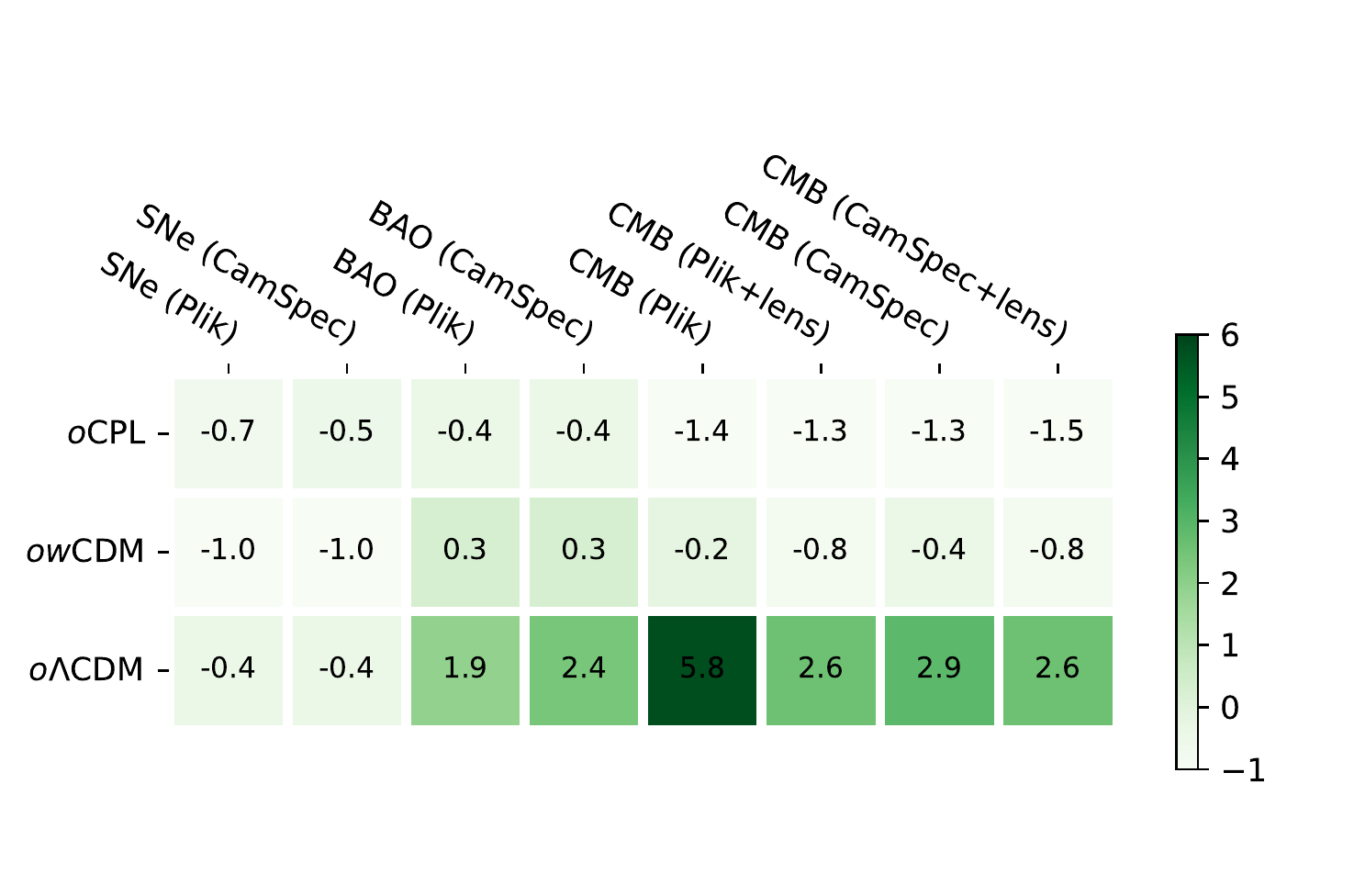}
\caption{Values of the ${\mathcal{O}}_j$  values as a measure of the compatibility between the constraints from the $j$-th data set and the joint constraints from the rest of the data sets. For SNe and BAO data, the ${\mathcal{O}}_j$ value is calculated considering CMB constraints from the Plik and CamSpec likelihoods. The color scale represents the ${\mathcal{O}}_j$ value. }
\label{fig:outliers} 
\end{figure*}

\section{Quantitative tension estimates}

In what follows, we quantify the results of the previous section and estimate possible tensions between the analysed data sets taking into account the full posterior parameter distributions and using the Index of Inconsistency as a discordance estimator, and the level of physical inconsistency as a tension measure caused by physical effects~\cite{Ishak2017a,Ishak2017b, Ishak2020} (see also Sec. \ref{1}).

Fig. \ref{fig:tensions} presents  heatmaps with the  $\beta$ values of pairs for data sets. The left, middle and right maps correspond to the $\Lambda$CDM, $w$CDM and CPL models, respectively, where blue maps represent spatially flat models while green maps represent models with non-zero curvature. For the $\Lambda$CDM model, we note that all data are consistent, i.e., no tension exceeds the threshold of evidence of physical inconsistency $P=0.85$ (see Eq. \ref{threshold}), ranking (1)  of the Table \ref{scale}, and the final result is a high precision estimate of the six model parameters. For the $w$CDM and $o\Lambda$CDM models, we note that the tensions increase, being higher for BAO and CMB(Plik) data, which shows substantial and substantial-to-strong inconsistencies, respectively.  The use of  the CamSpec likelihood alleviates the inconsistencies, but it still remains in the {{substantial}} rank for the $o\Lambda$CDM model. In this heatmap it is also seen that the addition of  the lensing power spectrum  to the CMB data  alleviates the physical inconsistencies, and we obtain non-substantial tensions.

Extending our analysis to the $ow$CDM and CPL models, we find that the inconsistencies decrease when a two-parameter extension of the $\Lambda$CDM model is considered. It is worth noticing that for the $ow$CDM model,  substantial-to-strong inconsistency appears between CMB(Plik) and BAO data, although such inconsistencies become insignificant when the CMB lensing data are also considered.  Therefore, assuming the  $ow$CDM model and taking into account the lensing data, all discrepancies are in an acceptable level to perform joint analyses involving these three observables. The last model analysed, the $o$CPL model, does not exhibit considerable improvements in terms of the tension analysis. We also note that independently of the cosmological model adopted, the highest tensions  are always found for the BAO data\footnote{It is worth mentioning that the usual approach to measure BAO in galaxy surveys make use of a fiducial cosmology (usually the flat $\Lambda$CDM model) to transform observed redshifts and angles to the estimated angular and radial BAO peak positions. In \cite{Carter:2019ulk}, the impact of the fiducial model on the inferred BAO scale was discussed by considering flat $w$CDM models. The influence of a non-zero curvature is unknown.}.

In Fig. \ref{fig:outliers}, we show the calculations of the Outlier Index ${\mathcal{O}}_j$ considering  SNe, BAO and CMB+Lens data and models with non-zero curvature ($o\Lambda$CDM, $ow$CDM and $o$CPL). The heatmap  evaluates if one specific data set constitutes  an outlier with respect to the other two data sets, and in parenthesis we specify which CMB likelihood is used in the analysis. In the case of the Outlier Index for the CMB data, we also calculate it  without considering the CMB lensing (fifth and seventh column). As can be seen, the unique outlier found with high significance is the CMB when the high multipoles of the temperature power spectrum are analyzed with the Plik likelihood.  The same significance is not obtained for CamSpec likelihood or when the lensing power spectrum measurements are taken into account.

\section{Cosmic Chronometers}

Recently, the authors of \cite{Vagnozzi2020b} argued that measurements of the expansion rate $H(z)$ from the relative ages of passively evolving galaxies \cite{Jimenez:2001gg,Simon:2004tf,Stern:2009ep,Moresco:2015cya,Zhang:2012mp,Moresco:2016mzx,Ratsimbazafy:2017vga} -- cosmic chronometers (CC) -- can break the geometrical degeneracy discussed above without presenting tensions with the remaining data sets.

\begin{figure}[t!]
\centering
\includegraphics[width=0.48\textwidth]{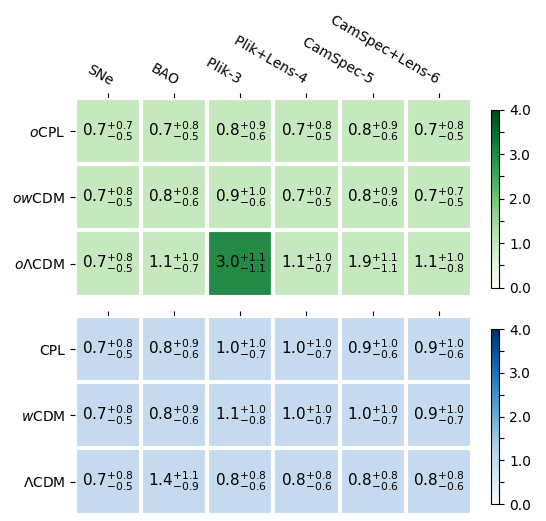}
\caption{Values of the level of physical inconsistency $\beta$ between CC and the other independent data sets assuming a flat universe (blue) and a universe with curvature (green). The color scale represents the ranking number introduced in Table \ref{scale}. Note that the highest inconsistency rank obtained is substantial-to-strong (3).}
\label{fig:tensionsCC} 
\end{figure}

By comparing the constraints obtained on the cosmological parameters from the joint analysis of CMB(Plik)+BAO in the framework of the $o\Lambda$CDM model~\cite{Vagnozzi2020a,Aghanim:2018oex}, 
\begin{eqnarray*}
H_0 & = & 67.88\pm0.66\;  \rm{km.s^{-1}.Mpc^{-1}}\\
\Omega_m & = & 0.310\pm 0.007  \\
\Omega_K & = & 0.0008 \pm 0.0019 
\end{eqnarray*}  with the constraints obtained from the joint analysis of CMB(Plik)+CC \cite{Vagnozzi2020b}, 
 \begin{eqnarray*}
H_0 & = & 65.23\pm 2.14\; \rm{km.s^{-1}.Mpc^{-1}}\\
\Omega_m & = & 0.336\pm 0.022\\
\Omega_K & = & -0.0054\pm 0.0055, 
\end{eqnarray*} 
it is possible to infer that (i) the results from CC and BAO data are not inconsistent and,  (ii) given the results presented in Fig.~(\ref{fig:LCDM}), there seems to be a tension between CMB (Plik) and CC data. 

In Fig. \ref{fig:tensionsCC}, we show the heatmaps of  the level of physical inconsistency $\beta$ between the CC and the other data sets, considering non-flat   (green) and flat (blue)  cosmologies. As seen in the upper panel of Fig. \ref{fig:tensionsCC}, we find a substantial-to-strong inconsistency between CMB(Plik) and CC data, with $\beta = 3.0^{+1.1}_{-1.1}$. This result changes substantially when the lensing power spectrum measurements are taken into account ($\beta = 1.1^{+1.0}_{-0.7}$), and differs from the mild disagreement found in Ref. \cite{Vagnozzi2020b}, which uses the ${\mathcal{I}}$ diagnostic based on the deviance information criterion (DIC). Finally, we also note that the tension between CC and CMB observations is significantly reduced when the CamSpec likelihood is used. Moreover, we find no significant inconsistency involving the CC data when a flat geometry is assumed in the analysis (lower panel of Fig.~\ref{fig:tensionsCC}).

\section{Conclusions}
\label{3}

Some of the most important aspects of our present understanding of the cosmic evolution are based on the results of joint analyses involving the currently available observational data. In this paper we assessed the reliability of such analyses by quantifying possible inconsistencies between the data sets through the Index of Inconsistency, the Outlier Index and the level of physical inconsistency $\beta$,  defined in Sec. IIb.  We used the latest observations of CMB, SNe Ia, BAO and CC and discussed the actual dependence of the results of their combined analysis with the spatial curvature parameter and dark energy parameterizations.

We presented the calculations of the level of physical inconsistency $\beta$ between all independent data sets considering six cosmological models mentioned in Sec. III.  We found that a model extension that considers a non zero curvature is necessary to describe all the features present in the data sets (separately) considered in this work, specially for the Planck CMB data (temperature+polarisation). This is  evident when the $\Lambda$CDM model is extended to the $o\Lambda$CDM and substantial-to-strong tensions appear between BAO and CMB data sets. However, our analysis also confirms that by considering jointly all data sets or even only CMB Planck information (temperature+polarisation+lensing), there is no  significant evidence for a non-flat Universe. 

As shown in Figs. \ref{fig:tensions} and \ref{fig:outliers}, the $ow$CDM is the simplest model showing only weak inconsistencies between all cosmological data including the CMB lensing information. Nevertheless,  when the joint analysis of CMB+Lens+SNe+BAO is performed, the curvature and EoS parameters are tightly constrained, i.e., 
$\Omega_k  = -0.0001^{+0.0023}_{-0.0021}$ and 
$w=-1.026^{ +0.039}_{ -0.032}$, 
and the model is reduced to the flat $\Lambda$CDM cosmology\footnote{ As shown in Ref. \cite{Alam:2020sor}, an analogous  result is also obtained for the $o$CPL model where the flat $\Lambda$CDM cosmology is recovered within $\sim 1\sigma$ CL by considering Pantheon, CMB(Plik), SDSS BAO+RSD and DES cosmic shear, galaxy  clustering,  and  galaxy-galaxy  lensing  data. }. On the other hand, we found no considerable improvement of the data tensions when an evolving dark energy EoS  is allowed, which means that considering higher order extensions of the standard cosmology (such as the $o$CPL model) do not seem to be the way to alleviate data inconsistencies. 

Finally, our analysis did not find significant tensions that a prevent joint analysis of the currently available cosmological data in non zero curvature models when lensing data is considered. Therefore, the most relevant tension found is the one between the CMB(Plik) and BAO data sets, which can possibly be solved with  the data from the next generation of galaxy surveys such as J-PAS, Euclid and DESI.

\acknowledgments

We thank Gabriela Antunes, Weikang Lin and Mustapha Ishak for very useful discussions. JEG acknowledges the Federal University of Rio Grande do Norte for financial support during my lockdown in Natal. MB acknowledges the Istituto Nazionale di Fisica Nucleare (INFN), sezione di Napoli, iniziativa specifica QGSKY.  RvM acknowledges financial support from the Programa de Capacita\c{c}\~ao Institucional (PCI) do Observat\'orio Nacional/MCTI. JA is supported by the Conselho Nacional de Desenvolvimento Cient\'{\i}fico e Tecnol\'ogico CNPq (Grants no. 310790/2014-0 and 400471/2014-0) and Funda\c{c}\~ao de Amparo \`a Pesquisa do Estado do Rio de Janeiro FAPERJ (grant no. 233906). The computational analyses of this work were developed thanks to the High Performance Computing Center at the Federal University of Rio Grande do Norte (NPAD/UFRN) and the National Observatory Data Center (DCON).

\bibliography{main}

\end{document}

%% file: preamble.tex
\usepackage{amsthm}
\usepackage{mathtools}
\usepackage{physics}
\usepackage{xcolor}
\usepackage{graphicx}
\usepackage[left=23mm,right=13mm,top=35mm,columnsep=15pt]{geometry} 
\usepackage{adjustbox}
\usepackage{placeins}
\usepackage[T1]{fontenc}
\usepackage{lipsum}
\usepackage{csquotes}